\documentclass[epj]{webofc}
\usepackage[utf8]{inputenc}
\usepackage[varg]{txfonts}   
\usepackage{booktabs}
\usepackage{xcolor}
\definecolor{darkred}{rgb}{0.4,0.0,0.0}
\definecolor{darkgreen}{rgb}{0.0,0.4,0.0}
\definecolor{darkblue}{rgb}{0.0,0.0,0.4}
\usepackage[bookmarks,linktocpage,colorlinks,
    linkcolor = darkred,
    urlcolor  = darkblue,
    citecolor = darkgreen]{hyperref}
\wocname{EPJ Web of Conferences}
\woctitle{Lattice2017}
\newcommand{\be}{\begin{equation}}
\newcommand{\ee}{\end{equation}}
\newcommand{\bea}{\begin{eqnarray}}
\newcommand{\eea}{\end{eqnarray}}

\newcommand{\st}{\sqrt{t_0}}

\begin{document}
%
\selectlanguage{english}
\title{%
Spectroscopy of the BSM sextet model
}
\author{%
\firstname{Zoltan} \lastname{Fodor}\inst{1,2,3} \and
\firstname{Kieran} \lastname{Holland}\inst{4} \and
\firstname{Julius}  \lastname{Kuti}\inst{5} \and
\firstname{Daniel}  \lastname{Nogradi}\inst{3,6,7} \and
\firstname{Chik Him}  \lastname{Wong}\inst{1}\fnsep\thanks{Speaker}
}
\institute{%
University of Wuppertal, Department of Physics, Wuppertal D-42097, Germany
\and
Juelich Supercomputing Center, Forschungszentrum Juelich, Juelich D-52425, Germany
\and
Eotvos University, Pazmany Peter setany 1/a, 1117 Budapest, Hungary
\and
University of the Pacific, 3601 Pacific Ave, Stockton CA 95211, USA
\and
University of California, San Diego, 9500 Gilman Drive, La Jolla CA 92093, USA
\and
MTA-ELTE Lendulet Lattice Gauge Theory Research Group, Budapest 1117, Hungary
\and 
Universidad Autonoma, IFT UAM/CSIC and Departamento de Fisica Teorica,
28049 Madrid, Spain
}
\abstract{%
As part of our ongoing lattice study of ${\mathrm {SU(3)}}$ gauge theory with two flavors of fermions in the two-index symmetric representation (the sextet model), we present the current status of the  pseudoscalar particle spectrum. We use a mixed action approach based on the gradient flow to control lattice artifacts, allowing a simultaneous extrapolation to the chiral and continuum limits. We find strong evidence that the pseudoscalar is a Goldstone boson state, with spontaneously broken chiral symmetry and a non-zero Goldstone decay constant in the chiral limit. In agreement with our study of the gauge coupling $\beta$ function, we find the sextet model appears to be a near-conformal gauge theory and serves as a prototype of the composite Higgs BSM template. 
}
\maketitle
\section{Introduction}\label{intro}

If physics beyond the Standard Model (BSM) is realized by a strongly-interacting gauge theory with new fermionic degrees of freedom, one of the fundamental questions will be what kind of particle spectrum such a theory produces~\cite{Kuti:2014epa}. Rescaling QCD and the hadronic spectrum from the GeV to the Electroweak scale may be a poor approximation of the true spectrum, if one wants to know if e.g.~a specific model predicts a light Higgs-like composite scalar with all heavier particle states above $~\sim 2$~TeV. It is also important to study {\it ab initio} the particle spectrum of a proposed model using lattice simulations, since a theory may in fact be infrared conformal with no massive particle states in the massless fermion limit. Coupling such a theory to the Standard Model would not break electroweak symmetry in the standard way. A non-perturbative extraction of the particle spectrum should also be consistent with other non-perturbatively measured features of the same model, e.g.~the $\beta$ function of the gauge coupling, to bolster confidence that the theory's infrared behavior has been correctly identified.

We have been studying $\mathrm {SU(3)}$ gauge theory with two flavors of massless fermions in the two-index symmetric representation (the sextet model) as a promising BSM theory~\cite{Fodor:2009ar,Fodor:2012ty,Fodor:2015zna}. The model is economic with few new degrees of freedom and if spontaneous chiral symmetry breaking ($\chi$SB) occurs, the three Goldstone bosons exactly match the longitudinal modes of the Electroweak gauge bosons without excess. Our initial lattice studies of the particle spectrum indicate that $\chi$SB does take place~\cite{Fodor:2016pls,Fodor:2015vwa,Fodor:2012ty}. This complements our other non-perturbative work which shows that the $\beta$ function of the theory is small but non-zero, consistent with a walking theory~\cite{Fodor:2015zna}. We have continued our investigation of the particle spectrum, we present here an update for the Goldstone states and their chiral analysis.  

\begin{figure}[thb] 
  \centering
  \includegraphics[width=7cm,clip]{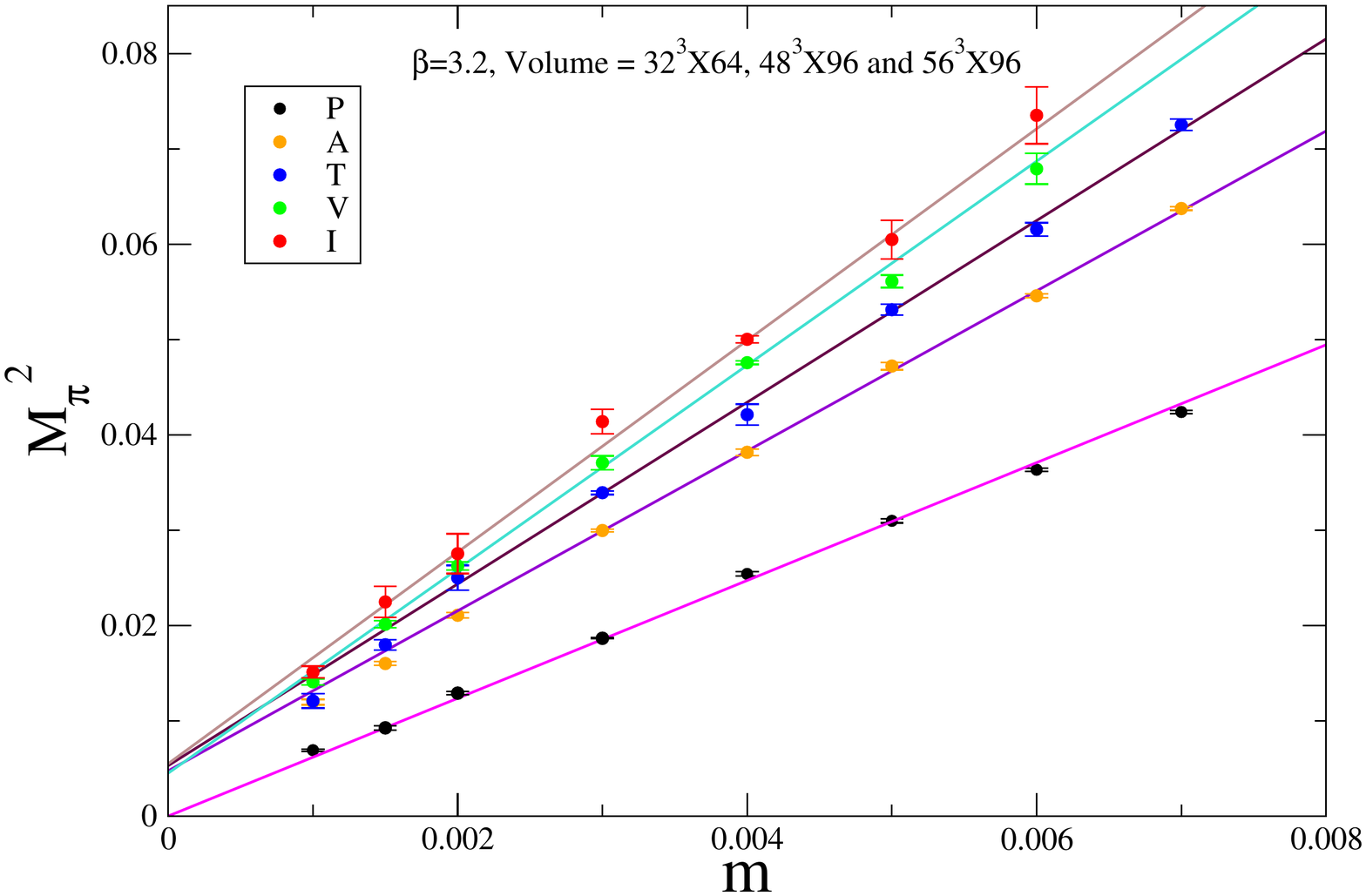}
  \includegraphics[width=7cm,clip]{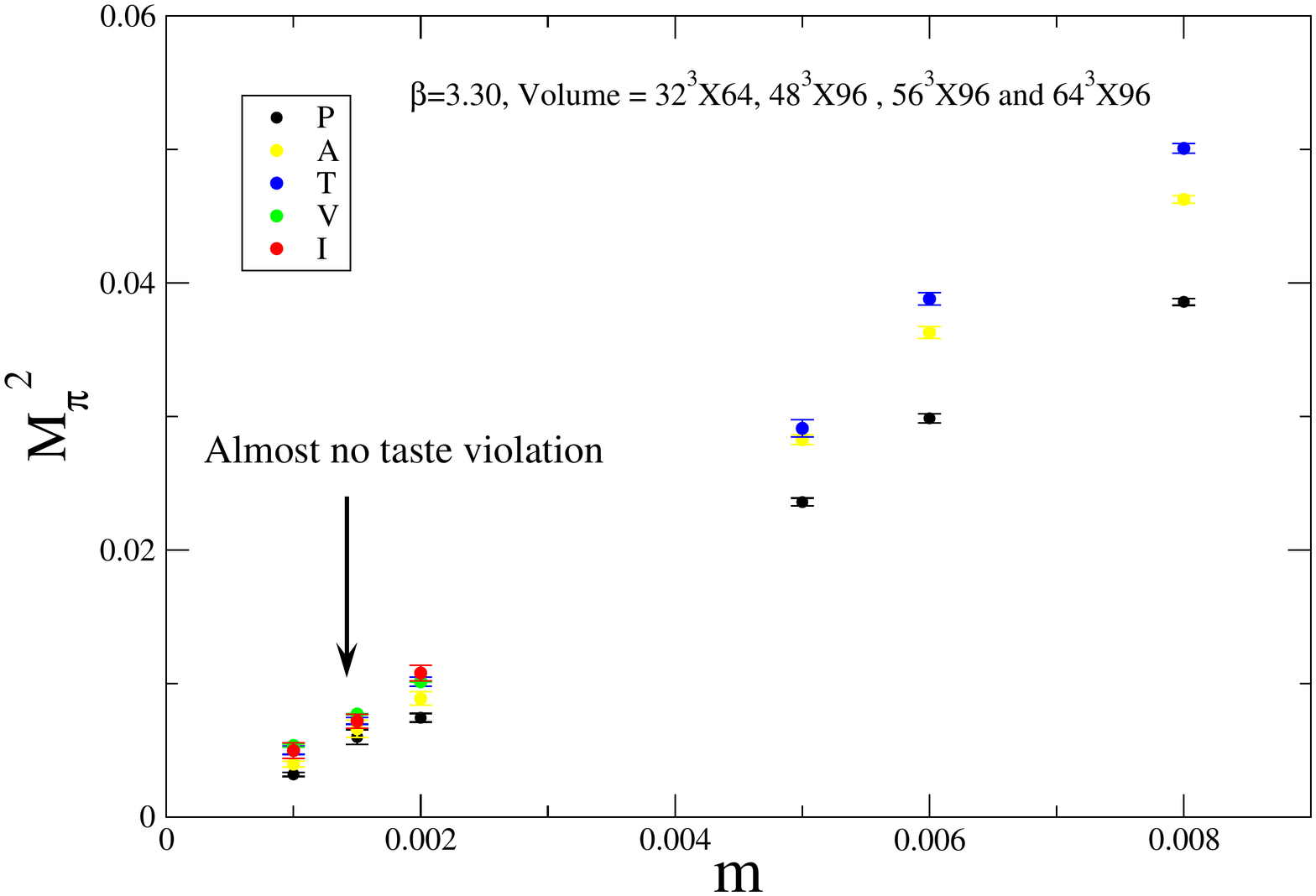}
  \caption{ Taste breaking in the Goldstone spectrum (left) at coarse lattice spacing corresponding to bare coupling $\beta = 3.20$ and (right) at fine lattice spacing with $\beta = 3.30$.}
  \label{fig-1}
\end{figure}

\section{Improvement}\label{improve}

We use staggered fermions with stout link improvement and the tree-level Symanzik improved gauge action as described in~\cite{Fodor:2012ty} in Monte Carlo simulations of the sextet model, for their computational speed and exact ${\mathrm {U(1)} }$ chiral symmetry, implementing the RHMC algorithm~\cite{Clark:2003na} to simulate two flavors. The staggered discretization introduces lattice artifacts in the particle spectrum, spoiling the degeneracy of certain states. We see in Figure~\ref{fig-1} the magnitude of this effect, known as taste breaking, in the Goldstone spectrum as the bare gauge coupling ranges from $\beta = 3.20$ to 3.30, which corresponds to factor of $\approx 1.6$ change in lattice spacing. The feature of unequal slopes for different tastes is unlike the analogous spectrum in staggered simulations of QCD, with parallel curves for different taste states. At finer lattice spacing with expected smaller artifacts, taste breaking is much reduced as the chiral limit is approached.

\begin{figure}[thb] 
  \centering
  \includegraphics[width=6.5cm,clip]{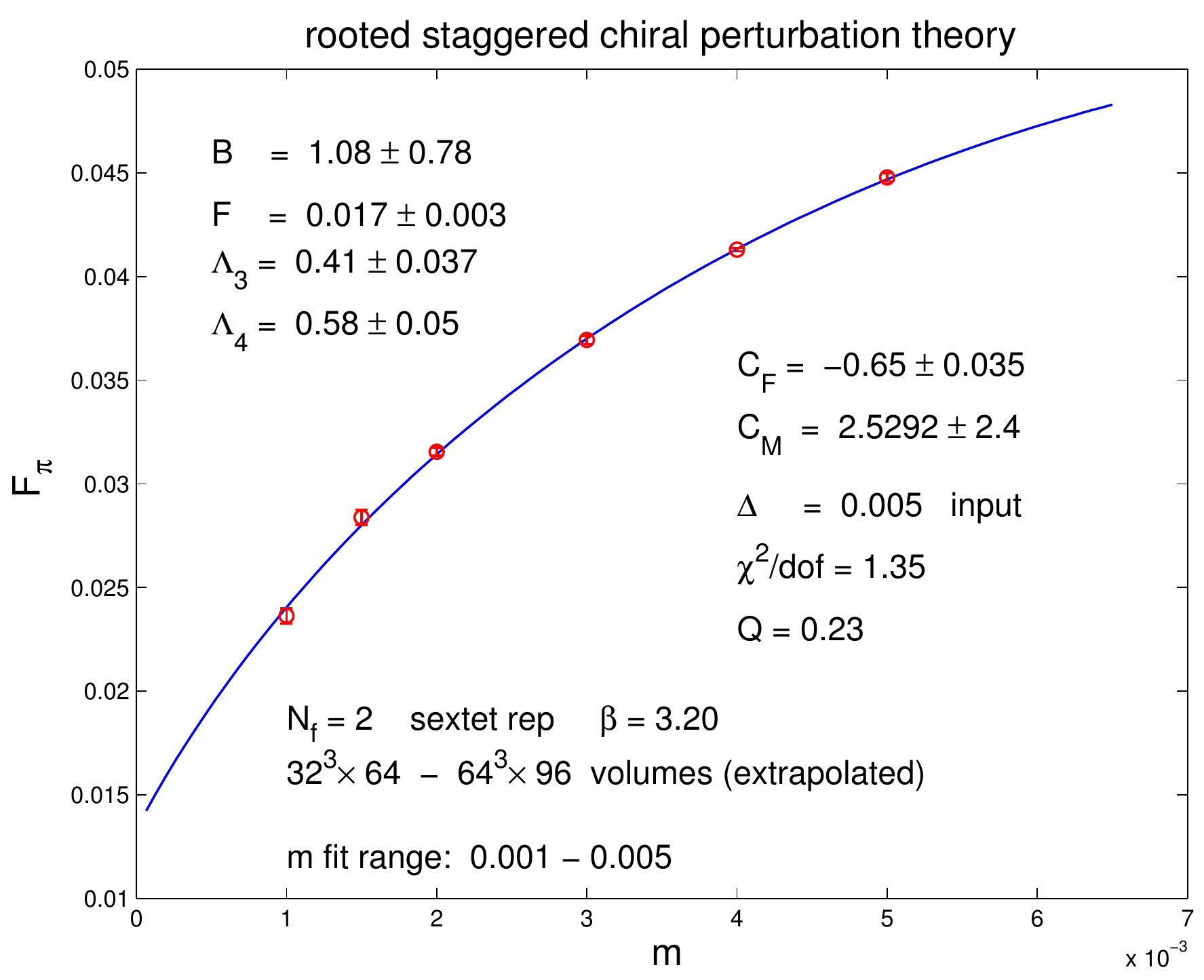}
  \includegraphics[width=6.5cm,clip]{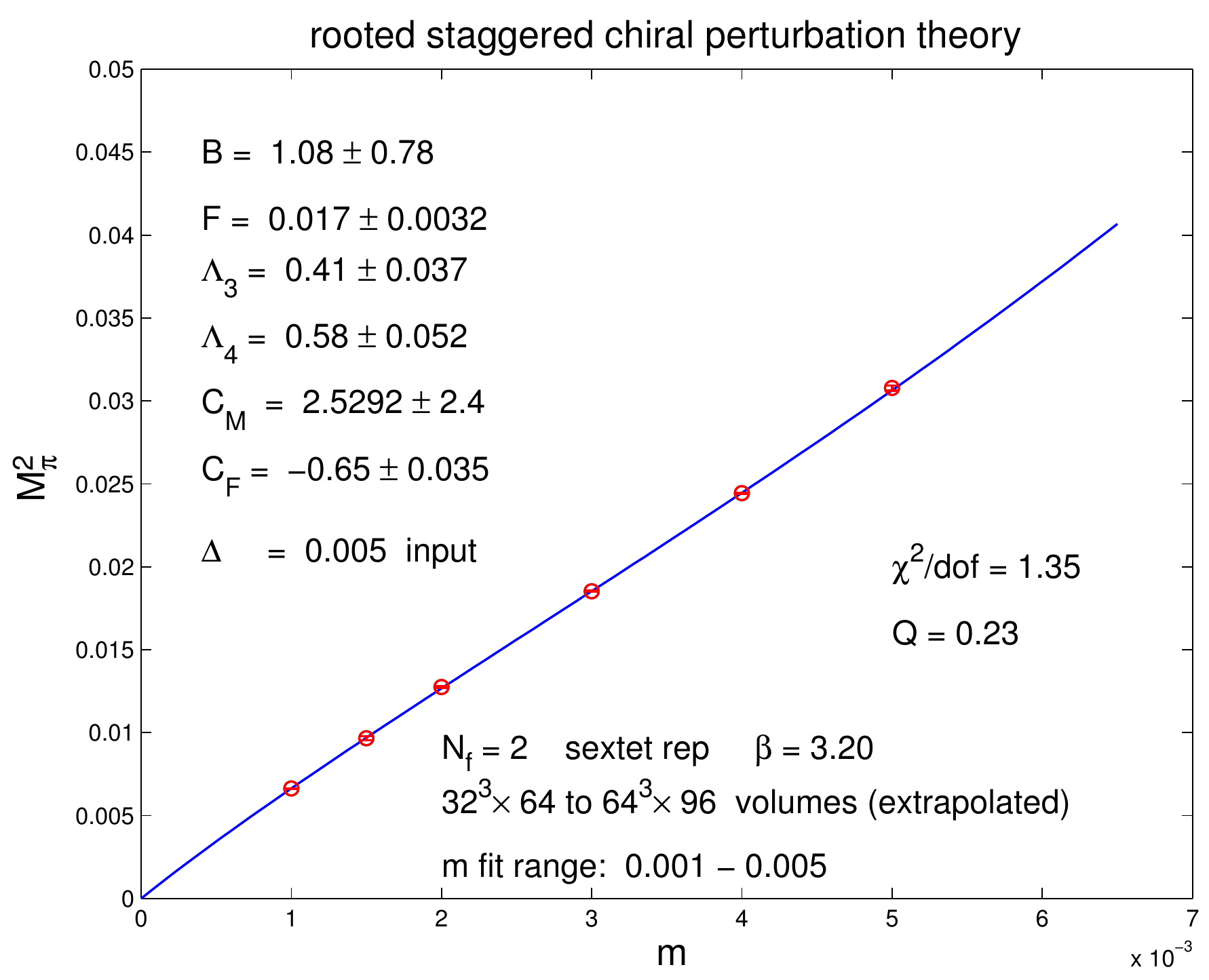}
  \caption{Rooted staggered chiral perturbation theory applied to (left) the Goldstone mass and (right) the Goldstone decay constant, both on the coarser lattice spacing at $\beta = 3.20$.}
  \label{fig-2}
\end{figure}

One approach to handling this lattice artifact is to use rooted staggered chiral perturbation theory~\cite{Aubin:2003mg} to extrapolate to the chiral limit, augmenting the theory to include taste-breaking terms. We show in Figure~\ref{fig-2} results at the coarser lattice spacing, fitting the fermion mass dependence of both the Goldstone boson mass and decay constant~\cite{Fodor:2016pls}. The method works well, however given the number of additional parameters, some assumptions have to be made, and the mass dependence of the decay constant is steep. The consistency of the results -- a vanishing Goldstone mass with a non-zero decay constant -- indicates that spontaneous $\chi$SB does occur. Our further investigation is to bolster this evidence as we go towards the continuum limit. 

Lattice artifacts which violate chiral symmetry can be restricted by using a numerically expensive chiral valence fermion, e.g.~overlap or domain wall fermions, to measure the particle spectrum on gauge configurations generated using inexpensive staggered sea fermions. Such a mixed action method has been successfully implemented in studying QCD~\cite{Beane:2011zm,Aubin:2008ie}. A numerically cheaper approach is to use the gradient flow~\cite{Luscher:2010iy,Narayanan:2006rf,Lohmayer:2011si,Luscher:2011bx} on the gauge configurations such that taste breaking using standard staggered fermions with the smoothed gauge links is much reduced. As long as the flow radius $\sqrt{8 t}$ is held fixed in lattice spacing units in the gradient-flow improved action, unitarity violations disappear from the continuum theory. An advantage of the flow improvement method is the computational ease of its implementation, where different flow times can be tested for the quality of restored taste symmetry. 

\begin{figure}[thb] 
  \centering
  \includegraphics[width=7cm,clip]{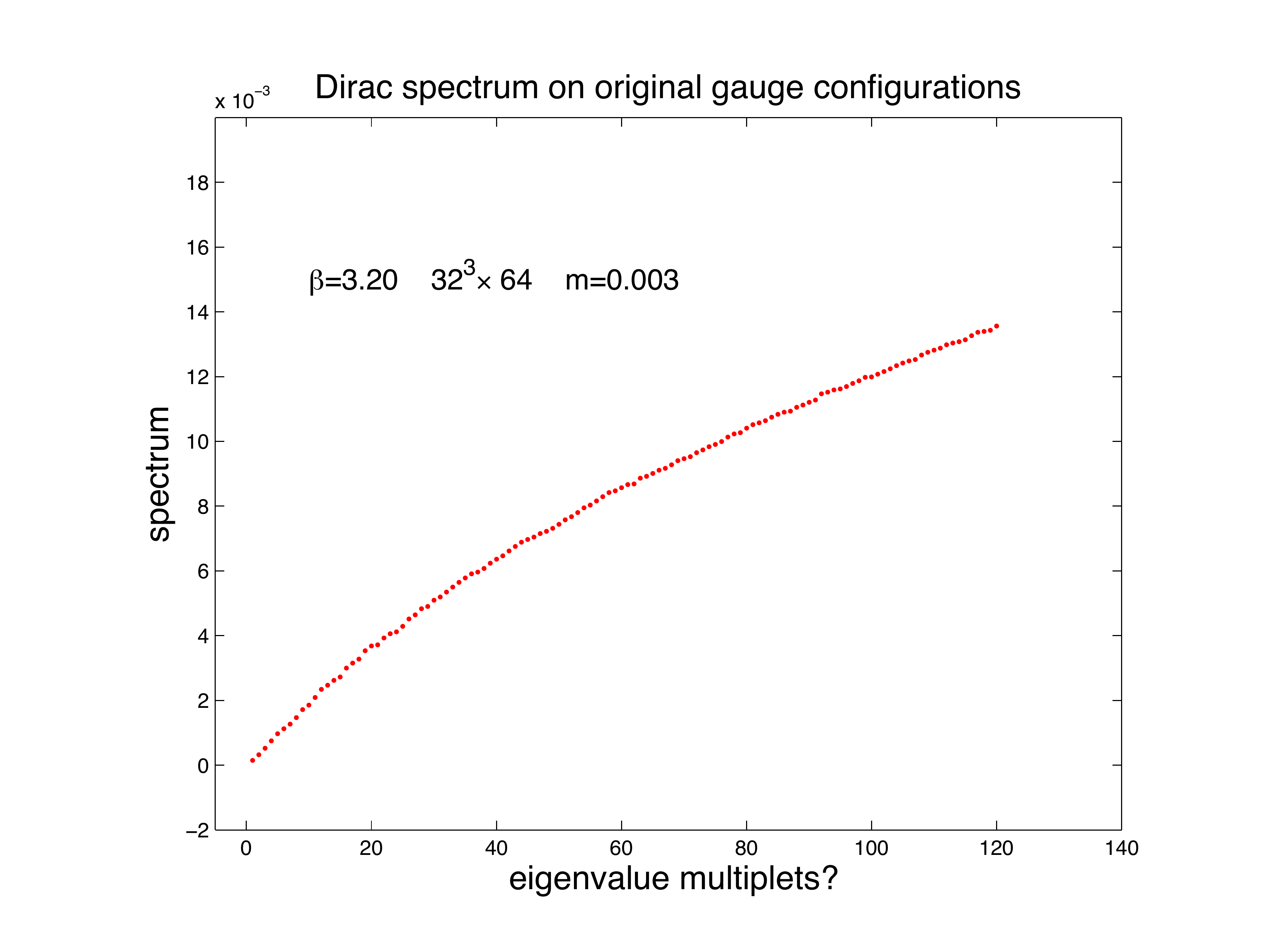}
  \includegraphics[width=7cm,clip]{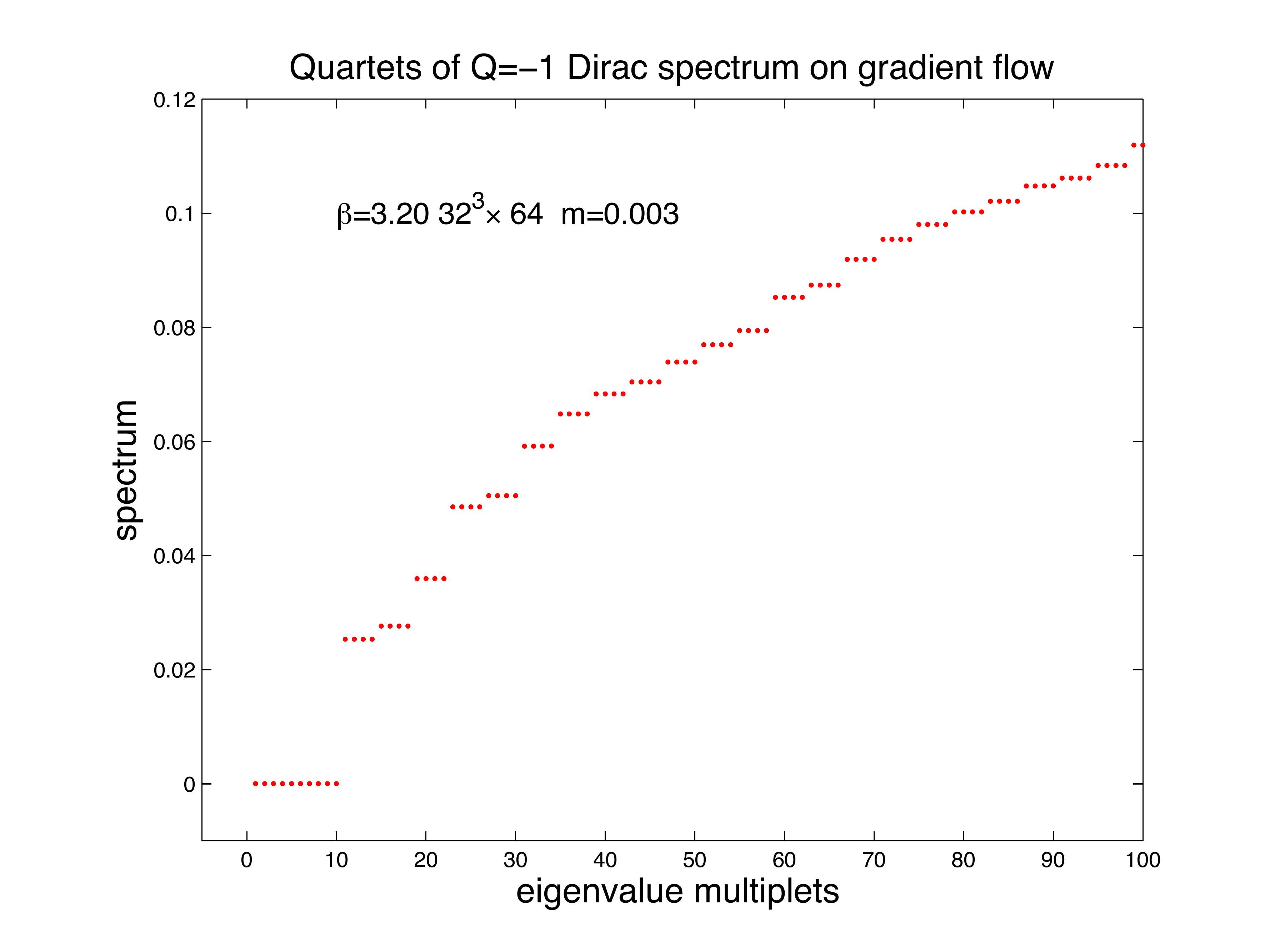}
  \caption{Eigenvalues of the staggered Dirac operator on gauge configurations with non-zero topology measured on (left) the original gauge configuration and (right) the smoothed gauge configuration where the gradient flow has been performed out to flow time $t/a^2=3$. The ten zero eigenvalues are a perfect match to the non-zero topology of the configuration, due to the multiplicity of the sextet representation in the index theorem.}
  \label{fig-3}
\end{figure}

\begin{figure}[thb] 
  \centering
  \includegraphics[width=7cm,clip]{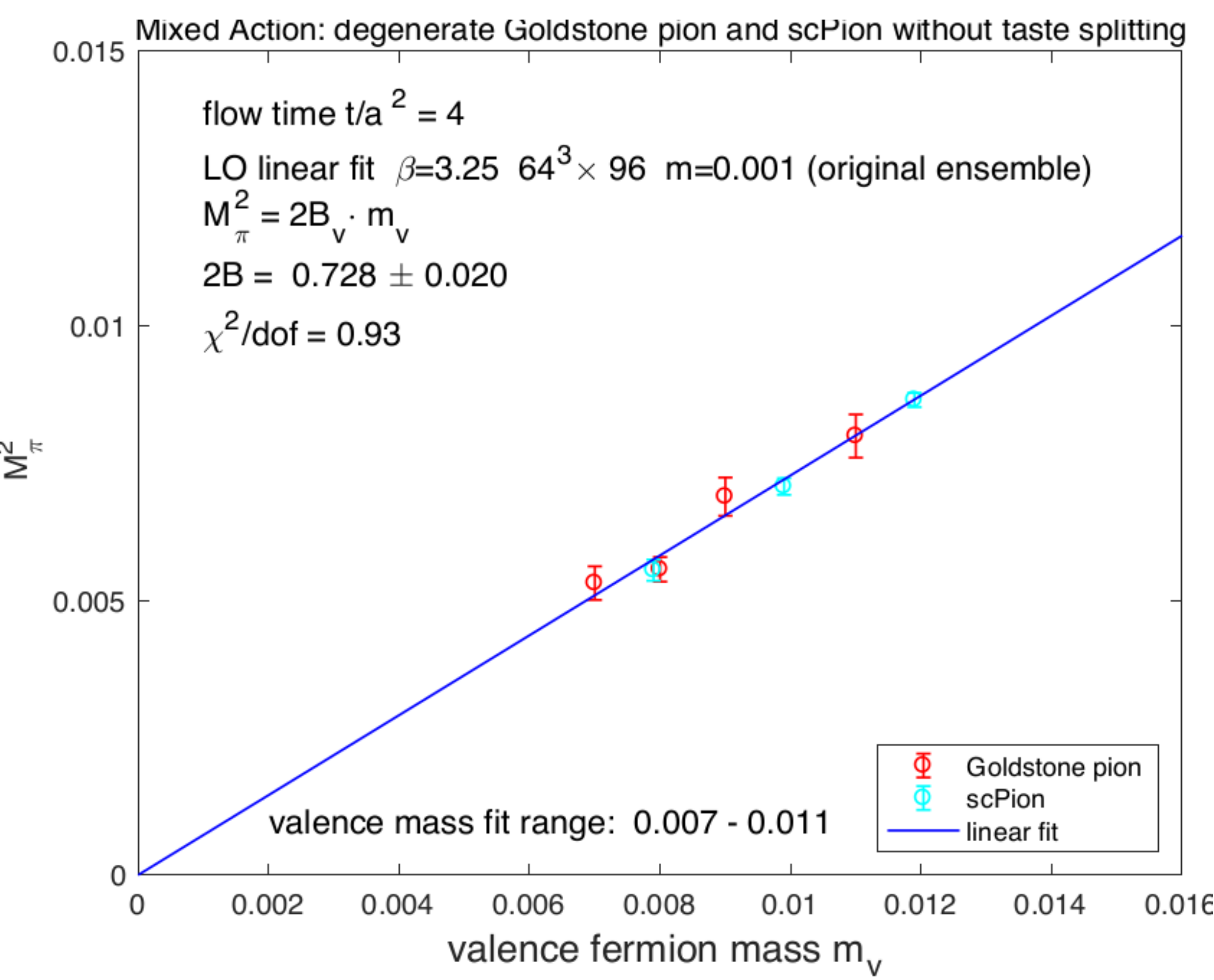}
  \includegraphics[width=7cm,clip]{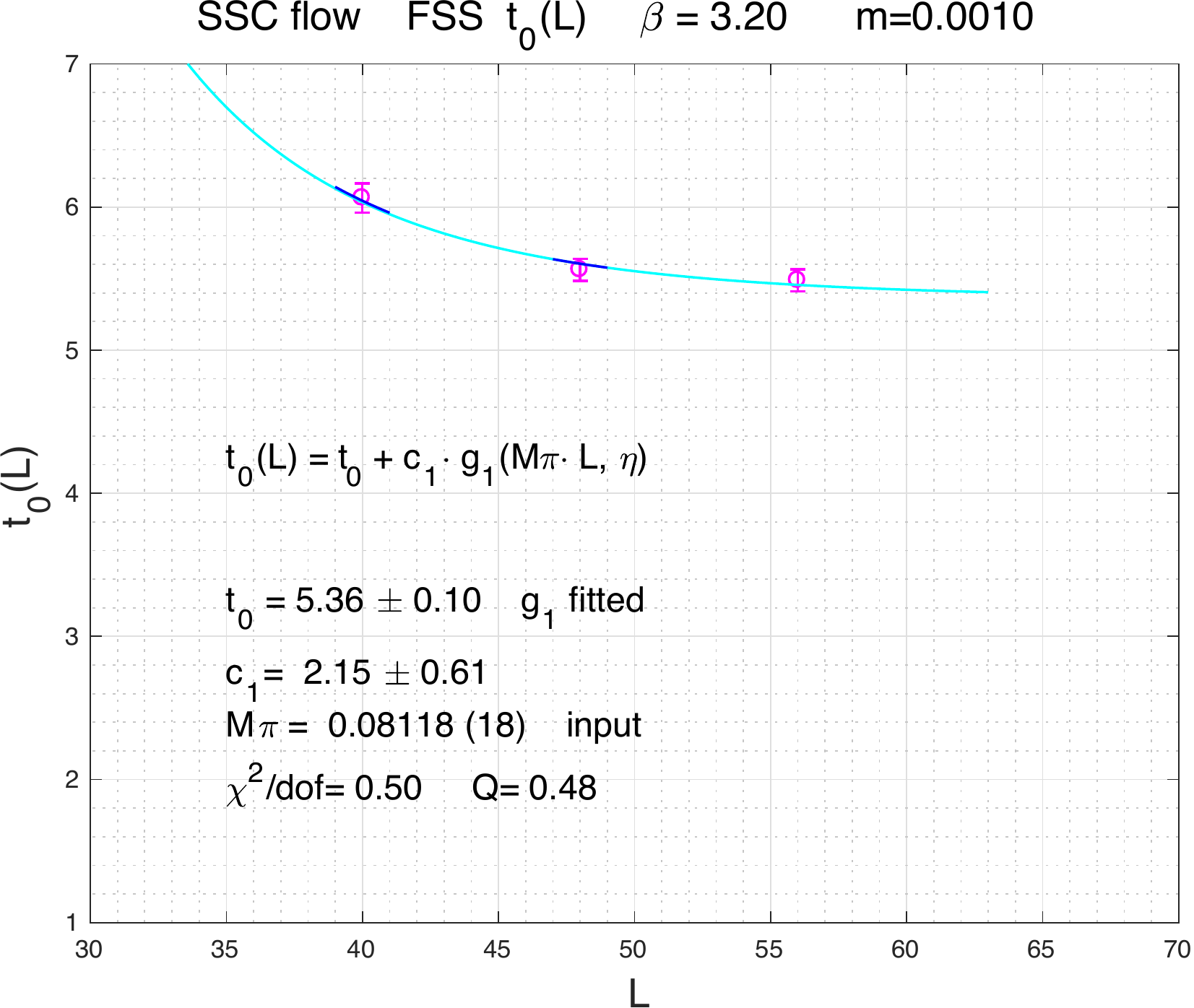}
  \caption{(left) Restoration of taste symmetry in the Goldstone spectrum using the gradient flow; (right) infinite volume extrapolation of the gradient flow scale $t_0$ at the coarsest lattice spacing and fixed fermion mass.}
  \label{fig-4}
\end{figure}

We show in Figure~\ref{fig-3} an example of the effect of the gradient flow on the eigenvalues of the staggered Dirac operator~\cite{Fodor:2015vwa}. On the initial Monte-Carlo-generated gauge configuration, there is no indication of quartet degeneracy in the eigenvalues or zero eigenvalues reflecting non-zero topology. These features emerge under the gradient flow, with the large multiplicity of zero eigenvalues due to the sextet representation in the index theorem~\cite{Fodor:2009ar,Fodor:2008hm}. The reduction of taste breaking by the flow is also observable in the Goldstone spectrum itself, shown in Figure~\ref{fig-4}, with different states having degenerate mass and leading order chiral perturbation theory describing the data well. 

The gradient flow can also be used to determine the lattice spacing in terms of a physical quantity. We do this via the choice of the renormalized gauge coupling i.e.~$g^2(t) \equiv (16 \pi^2/3) t^2 \langle E \rangle$ where the action density $E = (F_{\mu \nu}^a)^2/4$ expectation value is measured via simulations. The scale $t_0/a^2$ is set implicitly by the choice of renormalized coupling, we find $g^2(t_0) = 6.7$ is convenient for the lattice volumes and spacings we have used. (Note that this does not correspond to the choice of $t_0$ set by $t^2 \cdot \langle E \rangle_{t_0} = 0.3$ in the original investigation of~\cite{Luscher:2010iy}.) We will later use the scale $t_0$ to fit the ${\cal {O}}(a^2)$ lattice artifacts in the spectrum, similar to the fitting of lattice artifacts in the infinite volume $\beta$ function in~\cite{Fodor:lat17beta}. 

In Figure~\ref{fig-4} we show the volume dependence of $t_0$ at the coarser lattice spacing and smallest fermion mass. There are systematic effects when implementing the gradient flow, namely the discretization of the action density observable, and what gauge action is used to generate gauge configurations and in the gradient flow itself. We show the clover version of $E$ and the tree-level Symanzik-improved gauge action is used both in the Monte Carlo simulation and the gradient flow, hence the nomenclature SSC~\cite{Fodor:2014cpa}. We extrapolate to infinite volume using an ansatz with an infinite sum of Bessel functions dependent on the aspect ratio $L_t/L_s$ of the lattice volume to account for Goldstone bosons wrapping around the finite volume~\cite{Gasser:1986vb}~i.e.~$t_0(L) = t_0 + c_1 g_1(M_\pi L)$, in the numerical evaluation of $g_1$ the sum is truncated. The infinite volume Goldstone boson mass $M_\pi$ measured from the original configurations without flow is one of the input parameters for the fit --- this quantity is determined by an independent fit of $M_\pi(L)$ again using the $g_1$ fitting form. Once $t_0$ in the infinite volume limit is known, it is extrapolated to the chiral limit~\cite{Bar:2013ora} as
\be
t_0 = t_{\rm 0,ch} \left( 1 + k_1 \frac{M_\pi^2}{(4 \pi f)^2} + k_2 \frac{M_\pi^4}{(4 \pi f)^4} \log \left( \frac{M_\pi^2}{\mu^2} \right) + k_3 \frac{M_\pi^4}{(4 \pi f)^4} \right).
\label{eq1}
\ee
We find the leading-order linear behavior in $M_\pi^2$ describes the data at the lightest fermion masses well. We can alternatively extrapolate $t_0$ directly in the fermion mass $m$ itself, which should give consistent results to extrapolating in $M_\pi^2$. Our preliminary estimates for the scale $t_0/a^2$ are $6.20 \pm 0.14, 10.48 \pm 0.23$ and $15.85 \pm 0.46$ at the bare couplings $\beta = 3.20, 3.25$ and 3.30 respectively. An important consideration is that the gradient flow extent $\sqrt{8 t}/a$ should not outrun the Goldstone boson correlation length $1/(M_\pi \cdot a)$, otherwise the chiral expansion of the scale $t_0$ in Eq.~\ref{eq1} is invalid. 

\section{Chiral and continuum limits}\label{limits}

The measurement of the mixed action Goldstone boson mass and decay constant is straightforward: starting from the original gauge configurations, the gradient flow is carried out to a flow time fixed in lattice spacing $a$ units, the smoothed gauge configurations are then used for standard propagator calculation and contraction. A technical advantage of the flow is that the inversion of the Dirac operator converges more quickly than on the original gauge configurations due to the smoother underlying gauge links. In the following we hold the flow time fixed at $t/a^2 = 4$ at each lattice spacing in defining the mixed action.

\begin{figure}[thb] 
  \centering
  \includegraphics[width=6.9cm,clip]{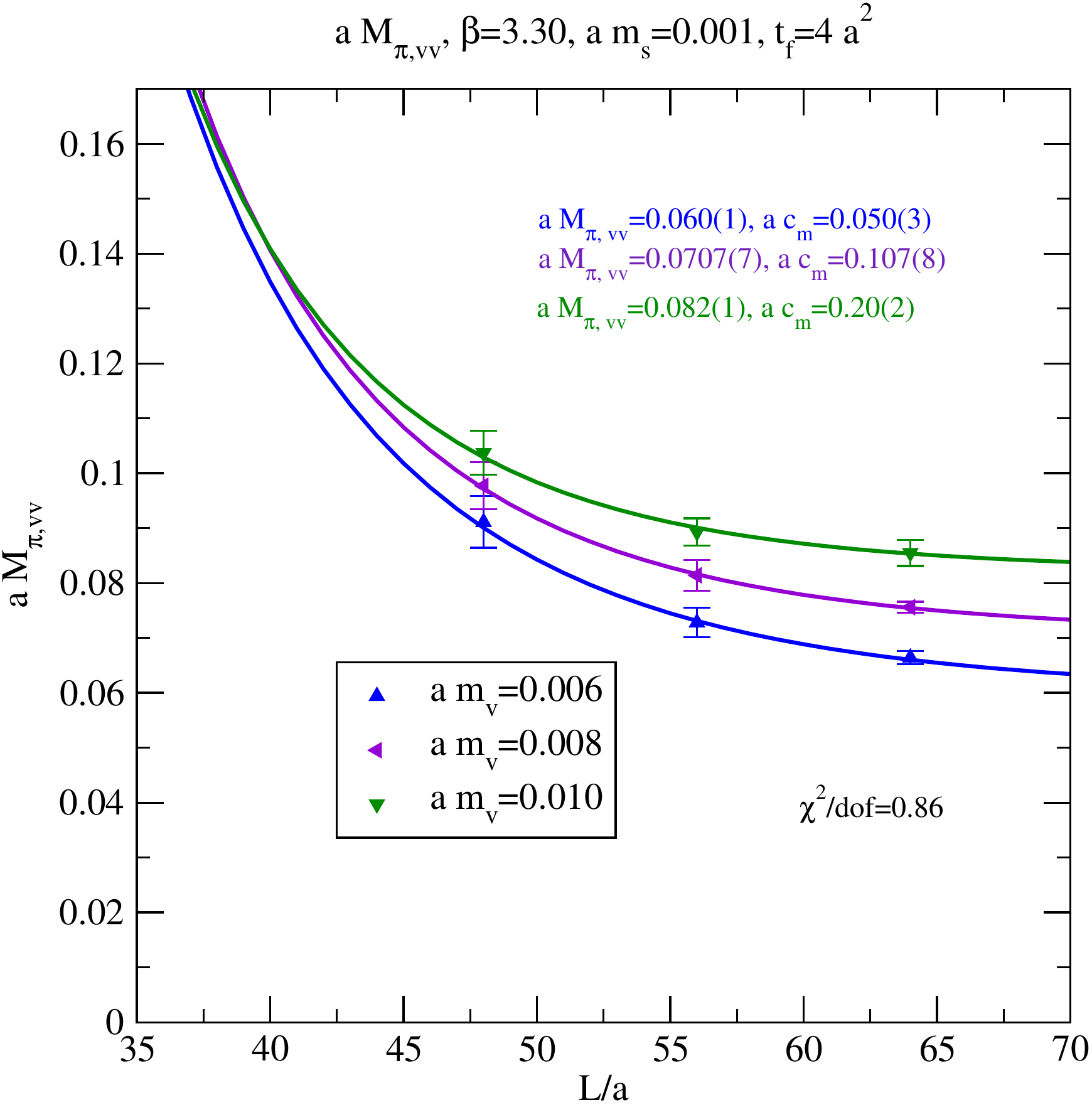}
  \includegraphics[width=7cm,clip]{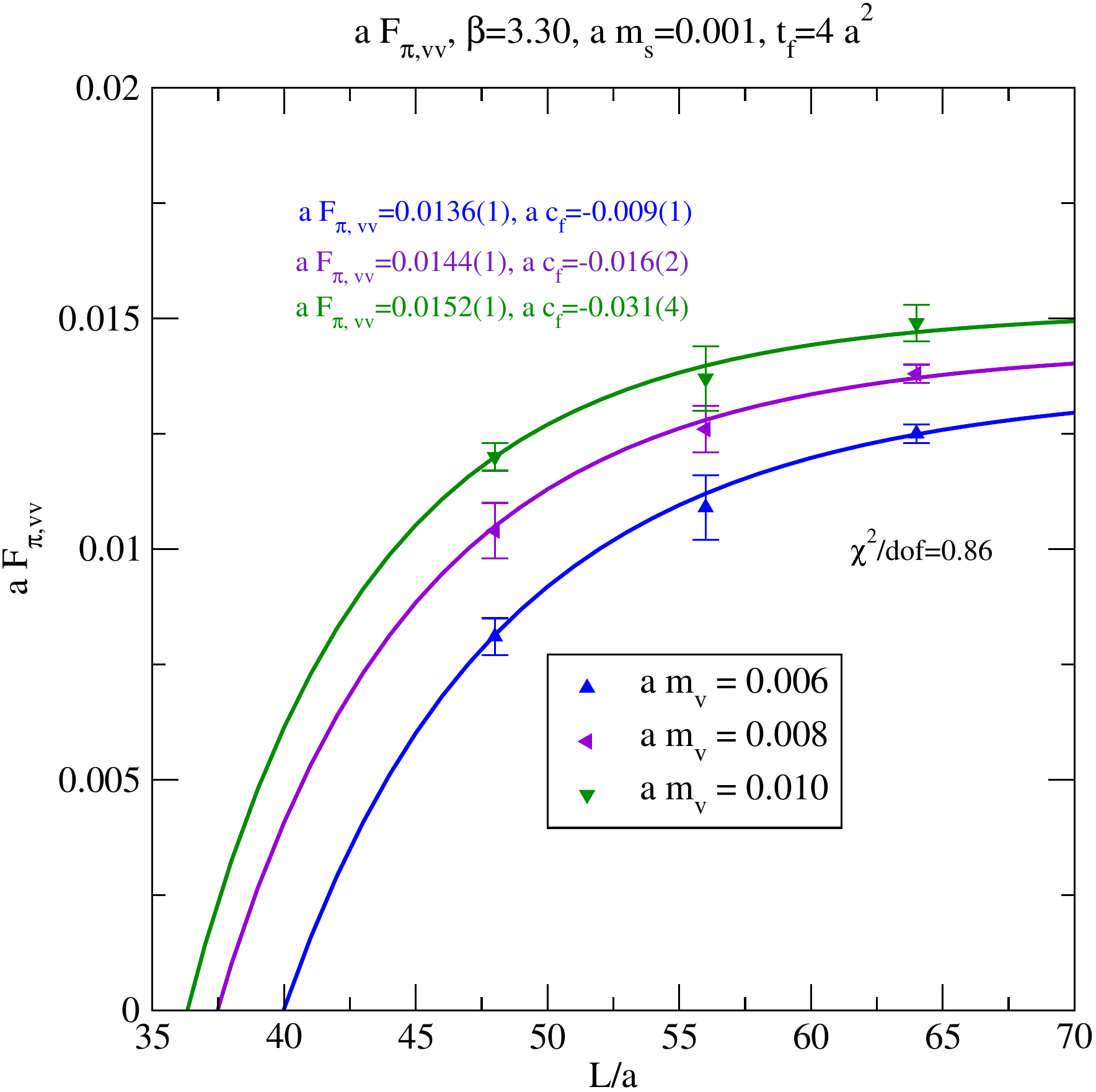}
  \caption{Infinite volume extrapolations of (left) the mixed action Goldstone boson mass and (right) the mixed action Goldstone decay constant for fixed sea fermion mass $a \cdot m_{\mathrm s} = 0.001$ and valence fermion masses $a \cdot m_{\mathrm v} = 0.006, 0.008$ and 0.010 at the finest lattice spacing.}
  \label{fig-5}
\end{figure}

As shown in Figure~\ref{fig-5} there can be noticeable finite volume dependence, which is sensitive to the choice of valence mass. The extrapolation to infinite volume uses the same fitting form of a sum of Bessel functions as used for the scale $t_0$. In the mixed action framework, the next-to-leading order (NLO) expressions for the mass dependence of the valence-valence Goldstone mass and decay constant are~\cite{Bar:2005tu}
\bea
M^2_{\pi,{\mathrm {vv,NLO}}} &=& M^2_{\pi,{\mathrm {vv,LO}}} \left( 1 + \frac{1}{32 \pi^2 F^2_\pi} \left[ ( M^2_{\pi,{\mathrm {vv,LO}}} - M^2_{\pi,{\mathrm {ss,LO}}} )  + 
(2 M^2_{\pi,{\mathrm {vv,LO}}} - M^2_{\pi,{\mathrm {ss,LO}}}) \ln \frac{M^2_{\pi,{\mathrm {vv,LO}}}}{\mu^2} \right] \right. \nonumber \\ 
&-& \left. \frac{8}{F^2_\pi} \left[ (L_5 - 2 L_8) M^2_{\pi,{\mathrm {vv,LO}}} + 2(L_4 - 2L_6) M^2_{\pi,{\mathrm {ss,LO}}} \right] \right) \nonumber \\
F_{\pi,{\mathrm {vv,NLO}}} &=& F \left( 1 - \frac{M^2_{\pi,{\mathrm {vs,LO}}}}{16 \pi^2 F^2} \ln \frac{M^2_{\pi,{\mathrm {vs,LO}}}}{\mu^2} + \frac{4}{F^2} \left[
L_5 M^2_{\pi,{\mathrm {vv,LO}}} + 2 L_4 M^2_{\pi,{\mathrm {ss,LO}}} \right] \right)
\label{eq4}
\eea
expressed in terms of the leading-order (LO) quantities $M^2_{\pi,{\mathrm {vv,LO}}} = 2B m_{\mathrm v}, M^2_{\pi,{\mathrm {ss,LO}}} = 2 B m_{\mathrm s}, M^2_{\pi,{\mathrm {vs,LO}}} = B( m_{\mathrm v} + m_{\mathrm s})$. Using the scale $t_0/a^2$ to re-express these in terms of dimensionless quantities gives
%
\bea
t_0 M^2_{\pi,{\mathrm {vv}}} &=&  \st m_{\mathrm v} \left[ 2B \st + b_1 \st m_{\mathrm v} + b_2 \st m_{\mathrm s} + b_3 \st (2 m_{\mathrm v} - m_{\mathrm s} ) \ln \st m_{\mathrm v} 
+ b_4 (a^2/t_0) \right] \nonumber \\
\st F_{\pi,{\mathrm {vv}}} &=& \st F + c_1 \st m_{\mathrm v} + c_2 \st m_{\mathrm s} + c_3 \st ( m_{\mathrm v} + m_{\mathrm s} ) \ln \st m_{\mathrm v} + c_4 (a^2/t_0),
\label{eq6}
\eea
where lattice artifact ${\cal O}(a^2)$ terms are added as the expected leading cutoff dependence. In this form the chiral and continuum extrapolations are performed together, yielding the physical parameters $2B$ (to be renormalized) and $F$ expressed in terms of the physical scale $t_0$.

\begin{figure}[thb] 
  \centering
  \includegraphics[width=6.9cm,clip]{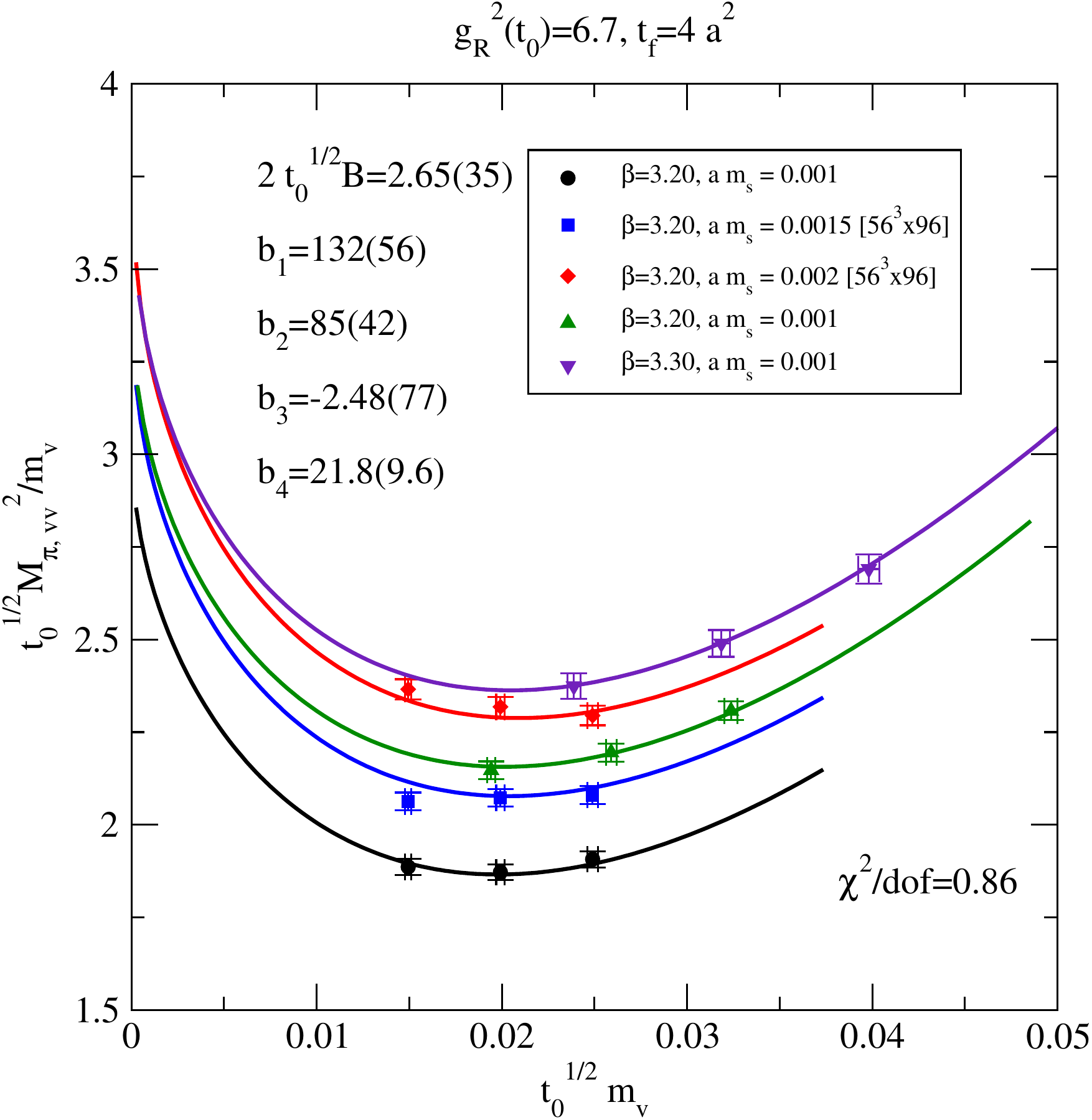}
  \includegraphics[width=7cm,clip]{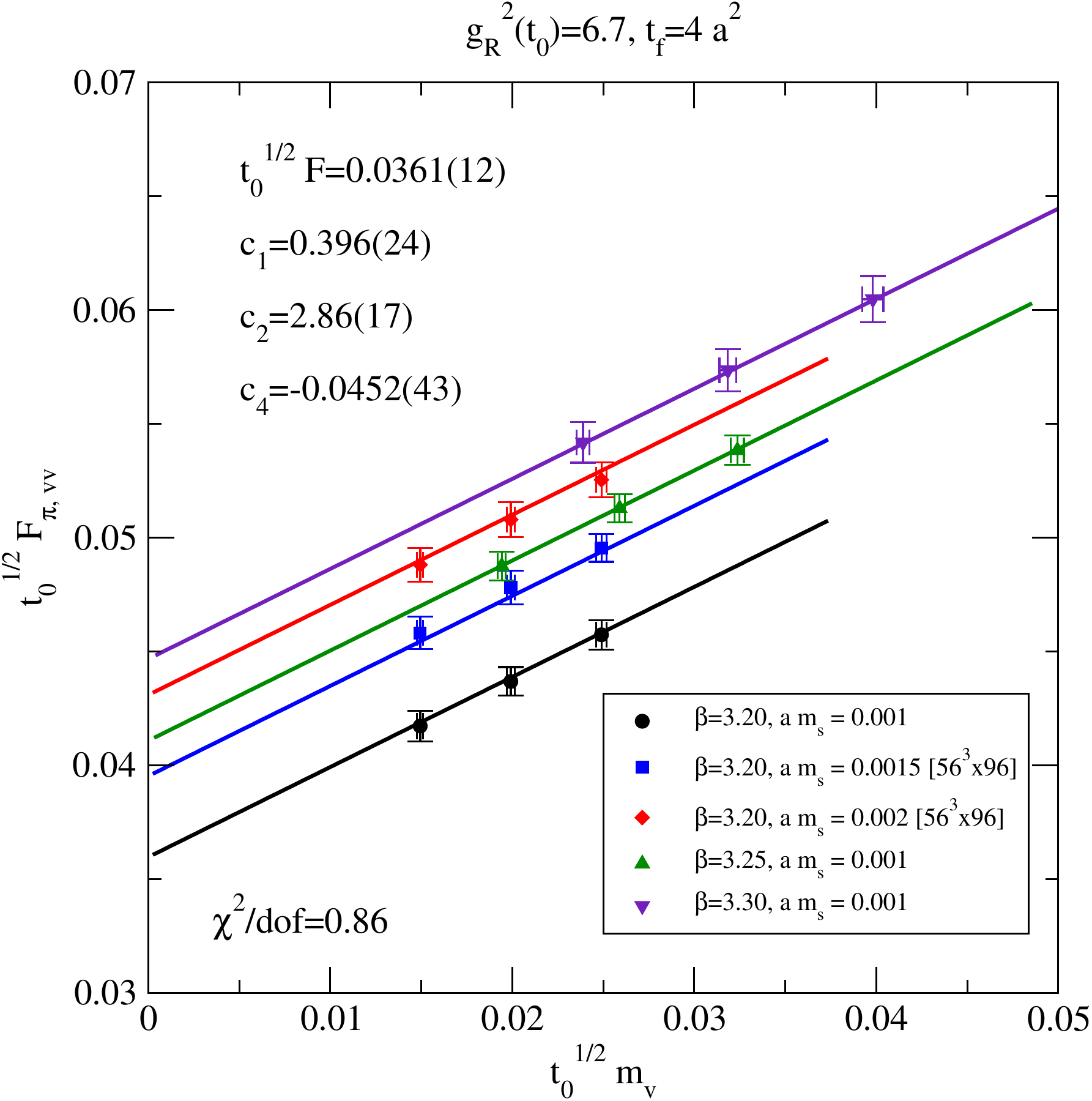}
  \caption{Combined chiral and continuum extrapolations of (left) the mixed action Goldstone boson mass and (right) the mixed action Goldstone decay constant. The parameters $b_1,...,b_4, c_1,...,c_4$ are defined in Eq.~\ref{eq6}.}
  \label{fig-6}
\end{figure}

One natural procedure to analyze the large set of data is to perform the infinite-volume extrapolations independently at each valence and sea fermion mass, followed by the simultaneous chiral-and-continuum fit of Eq.~\ref{eq6}. An alternative is to combine volume, chiral and continuum fits into one global fit i.e.~the sum of the residuals of volume-, mass- and cutoff-dependence is simultaneously minimized, hence there is one $\chi^2$/dof value for the entire fit. We show the results of the global fitting method in Figures~\ref{fig-5} and~\ref{fig-6}, where the overall quality of the fit is good. Plotting $t_0 M^2_{\pi,{\mathrm {vv}}}/\st m_{\mathrm v}$ amplifies the curvature in the data which is well-described by the chiral logarithm term in Eq.~\ref{eq6}. In contrast there is no indication of curvature in $\st F_{\pi,{\mathrm {vv}}}$ and the coefficient of the logarithmic term is consistent with zero within error. 

Using the other approach of first performing the infinite-volume fit, followed by the joint chiral and continuum fits, yields very similar results for the physical parameters. One can also use the value of $t_0$ at each finite sea mass $m_{\mathrm s}$, as opposed to its chiral limit value, in forming the dimensionless combinations $t_0 M^2_{\pi,{\mathrm {vv}}}, \st F_{\pi,{\mathrm {vv}}}, \st m_{\mathrm s}$ and $\st m_{\mathrm s}$. Although this mixes additional mass dependence with that of Eq.~\ref{eq6}, we find that this again gives very similar results for the parameters $2B$ and $F$. This is a valuable cross-check, as it avoids a chiral extrapolation for $t_0$ before the continuum extrapolation of $M^2_{\pi,{\mathrm {vv}}}$ and $F_{\pi,{\mathrm {vv}}}$, which removes any concern regarding the order of limits for staggered fermions. The error of $t_0$ appears both in the data to be fitted ($t_0 M^2_{\pi,{\mathrm {vv}}}$ and $\st F_{\pi,{\mathrm {vv}}}$) and in the fitting variables $\st m_{\mathrm v}, \st m_{\mathrm s}$ and $a^2/t_0$. We have tested the latter effect by generating Gaussian distributions of $t_0$ values based on the actual $t_0$ measurements and errors, then tracking the resulting spread of parameters from the fit according to Eq.~\ref{eq6}. We find the effect to be small compared to the statistical error of $M^2_{\pi,{\mathrm {vv}}}$ and $F_{\pi,{\mathrm {vv}}}$ themselves. In the chiral and continuum limit, there is no distinction between sea and valence mass, hence $2B$ (once renormalized) and $F$ are physical parameters. Their non-zero value is compelling evidence that the sextet model is non-conformal in the chiral limit.

\section{Comparison and conclusion}\label{conclude}

In the non-perturbative investigation of new models whose infrared behavior is to be determined, the control of systematic effects is crucial to be able to draw conclusions with confidence. Lattice artifacts are certainly one such important concern. Rooted staggered chiral perturbation theory at fixed lattice spacing describes the pseudoscalar mass dependence well, however it involves a larger set of parameters and some assumptions about the dominant sources of taste breaking. 

By taking the continuum limit of the Goldstone spectrum, along the way controlling the distortion due to working in finite volume, and accessing light enough fermion mass that a chiral extrapolation appears justified, we are able to state with more confidence that the particle spectrum supports the picture of spontaneous $\chi$SB. This places these results in the same light as our non-perturbative study of the finite-volume $\beta$ function via step-scaling~\cite{Fodor:2015zna}, whose continuum limit shows the model has a small $\beta$ function up to strong enough renormalized coupling to generate the particle spectrum, with a finite mass gap in the non-Goldstone states. Our spectrum results are in direct contradiction to those of~\cite{Hansen:2017ejh} using Wilson fermions, which prefer an infrared conformal interpretation of the spectrum of the sextet model. 

We have recently added a new independent study of the infinite-volume $\beta$ function, where the same scale $t_0$ is used as a measure of cutoff effects~\cite{Fodor:lat17beta}. Although they are different renormalization schemes, the infinite-volume $\beta$ function extracted from the chiral limit of simulations at finite fermion mass is in very good agreement with the finite-volume $\beta$ function where simulations are performed directly at zero fermion mass. Their consistency rebuts concerns about the use of staggered fermions and violations of universality in the study of the sextet model as stated in~\cite{Hasenfratz:2015ssa}.

Our previous work on the particle spectrum gives signs of a light composite scalar and indicates that the next lightest states occur above $\sim 2$~TeV~\cite{Fodor:2016pls}. It will be important to sharpen these predictions, obtained at finite lattice spacing, perhaps using a similar approach to what we present here. The mixed action chiral analysis apparently does not need additional terms to reflect the influence of a light scalar, which will require further investigation to confirm. We are also investigating if the light composite scalar shows dilaton-like features in extensions of chiral perturbation theory where spontaneous scale symmetry breaking is entangled with chiral symmetry breaking, with various possible forms of the dilaton potential~\cite{Kuti:lat17dila}. Further study will also be required to see if the dilaton potential might be directly measured via Monte Carlo simulations, whether the linear sigma model regime is attainable, and find possible signatures of a dilaton in the $\epsilon$- and $\delta$-regimes.

\section*{Acknowledgments}
We acknowledge support by the DOE under grant DE-SC0009919, by the NSF under grants 1318220 and 1620845, by OTKA under the grant OTKA-NF-104034, and by the Deutsche
Forschungsgemeinschaft grant SFB-TR 55. Computational resources were provided by the DOE INCITE program on the ALCF BG/Q platform, by USQCD at Fermilab, by the University of Wuppertal, by Juelich Supercomputing Center on Juqueen and by the Institute for Theoretical Physics, Eotvos University. We are grateful to Szabolcs Borsanyi for his code development for the BG/Q platform. We are also grateful to Sandor Katz and Kalman Szabo for their CUDA code development.

\bibliography{wong2017}

\end{document}